\documentclass{aa}  
\usepackage{graphicx}
\usepackage{longtable}
\usepackage{txfonts}
\begin{document}
\title{New high-sensitivity, milliarcsecond
resolution results from routine observations of lunar occultations
at the ESO VLT
\thanks{Based on observations made with ESO telescopes at Paranal Observatory}
}

\titlerunning{New 
results from observations of lunar occultations
at the ESO VLT
}

   \author{A. Richichi\inst{1}
          \and
          O. Fors\inst{2}\fnmsep\inst{3}
          \and
          W-P. Chen\inst{4}
          \and
          E. Mason\inst{5}
          }

   \offprints{A. Richichi}

   \institute{
             European Southern Observatory,
Karl-Schwarzschild-Str. 2, 85748 Garching bei M\"unchen, Germany
             \email{arichich@eso.org}
         \and
Departament Astronomia i Meteorologia and Institut de Ci\`encies 
del Cosmos (ICC),
Universitat de Barcelona (UB/IEEC), 
Mart\'{\i} i Franqu\'es 1, 08028 Barcelona, Spain
         \and
             Observatori Fabra, Cam\'{\i} de l'Observatori s/n, 08035 
Barcelona, Spain
         \and
Graduate Institute of Astronomy, National Central University,
 300 Jhongda Road, Jhongli 32054, Taiwan
   \and
             European Southern Observatory,
Santiago, Chile
             }

  \abstract
{
Lunar occultations (LO) are a very efficient and powerful technique,
that achieves the best combination of high angular resolution and
sensitivity possible today at near-infrared wavelengths.
Given that the events are fixed in time, that the sources
are occulted randomly, and that the telescope use is minimal,
the technique is very well suited for service mode observations. 
}
{
We have established a 
program of routine LO observations at the VLT observatory,
especially designed to take advantage of short breaks available
in-between other programs. We have used the ISAAC instrument
in burst mode, capable of producing continuous read-outs
at millisecond rates on a suitable subwindow.
Given the random nature of the source selection, our aim
has been primarily the investigation of a large number
of stellar sources at the highest angular resolution in
order to detect new binaries. 
Serendipitous results such as resolved sources and
detection of circumstellar components were also anticipated.
}
{
We have recorded the signal from background stars for
a few seconds, around the predicted time of occultation by the
Moon's dark limb. At millisecond time resolution, 
a characteristic diffraction
pattern can be observed. Patterns for two or more sources superimpose
linearly, and this property is used for the detection of binary stars. 
The detailed analysis of the diffraction
fringes can be used to measure specific properties such
as the stellar angular size and the presence of extended
light sources such as a circumstellar shell.
}
{We present a list of 191 stars for which LO data could
be recorded and analyzed. Results include the detection
of 16 binary and 2 triple stars, all but one of which 
were previously unknown. The projected
angular separations are as small as 4 milliarcseconds and
magnitude differences as high as $\Delta$K=5.8\,mag.
Additionally we derive accurate angular diameters for 2 stars
and resolve circumstellar emission around another one,
also all for the first time. 
We have
established upper limits on the angular size of 177
stars, mostly in the 1 to 5\,mas range, and we plan
to include them in a future list of sources well suited
for the calibration of interferometers.
}
{
We confirm the performance of the technique
already established in our previous work. LO 
at an 8\,m-class telescope can achieve an angular resolution close
to $0\farcs001$ with a sensitivity K$\approx 12$\,mag.
}

   \maketitle

\section{Introduction}
Lunar occultations (LO) are a simple and effective technique
to achieve high angular resolution far exceeding the diffraction
limit of any single telescope, and matched only by long-baseline
interferometry (LBI) which however has much more demanding
technical requirements. In a LO event, the light of a background
source occulted by the Moon falls to zero with a characteristic
diffraction pattern. For a simple monochromatic point-source,
the pattern is determined by the wavelength and by the distance
to the Moon. As the Moon moves over the source, the pattern
sweeps over a ground-based telescope with a speed which is 
typically of order $\approx 0.75$~m/ms, but which can be
significanly slower in case of events occurring with high
contact angles. The practical case of a resolved stellar disc
observed in a broad-band filter can be easily modelled
by the convolution of several point-like monochromatic
patterns. Likewise, smearing due to the telescope aperture and
to the finite detection time can also be modelled.
In the case of sources exceeding some characteristic angular
size, about 20 to 40 milliarcseconds (mas) at
near-infrared wavelengths, the diffraction fringes are 
completely smeared and the monotonic intensity fall intuitively
expected for geometrical optics is observed.

From these few considerations it is apparent that LO can be
used to measure stellar angular diameters. Moreover,
the diffraction patterns of two or more sources
add linearly, so LO can easily
detect binary and multiple stars. These two applications have
provided the bulk of LO results, starting from the first
photoelectic and infrared observations a few decades ago.
The CHARM2 catalog (Richichi et al. \cite{CHARM2}) lists
several hundreds of LO results both for angular diameters
and for binary stars.

A severe limitation of LO is of course the fact that
the sources are randomly selected in a restricted area
of the sky ($\approx 10$\%), and that they are fixed-time
events. As LBI progressed both in performance and in
reliability, in the past decade LO have gradually
lost their competitive edge. However, two important factors
are tipping the scale again in their favor: the availability
of sensitive infrared all-sky surveys, and the increase in
limiting magnitude provided by the use of very large telescopes.
In two previous papers
(Richichi et al. \cite{PaperI}, \cite{PaperII}, 
Paper~I and Paper~II hereafter) we have described the first
results from this renaissance in the LO technique, applied
to very crowded regions in the vicinity of the Galactic Center.
The present paper follows closely in the footsteps, and
for this reason we do not repeat here several of the
technical details already provided in Paper~I.
An important point however is 
the fact that now LO have been implemented as 
a routine service program at the ESO VLT. Here we present
the results from the first periods of service mode observation.

\section{Observations and Data Reduction}\label{data}
As before, we have used the burst mode available at the ISAAC
instrument. We have taken advantage of the fact that this
mode is now publicly offered and supported at ESO, and
we have proposed to observe LO as a filler between other
programs. In particular, given that LO can be satisfactorily
observed also under less than optimal sky conditions and
that each event requires only a relatively short telescope time,
we have aimed at filling gaps in which other service
programs could not be executed either because of weather
or because of the available slots being too short.

In each of three six-month periods 
starting  in October 2007, April 2008, and April 2009
(ESO P80, P81, P83 respectively)
we have been granted 15 hours of observing
time. In each period we have submitted so-called Observing 
Blocks (OB, i.e. templates for the automated telescope
and instrument operation) for about 1,000 LO events.
This sufficed to provide one event for any 5 minute
interval when the Moon was above the elevation limits,
when service mode was available, and
with the correct phase - LO cannot be observed on the
bright side of the limb and we restricted ourselves
to disappearances for service mode.
Our predictions were based on the 2MASS Catalog
(Cutri et al. \cite{2003yCat.2246....0C}), with a
limiting magnitude set to K=11, which is above our
nominal detection limit but already provides significantly
more events than can be practically observed.
Sources were then prioritized by their
brightness and colors, and when available 
on the basis of cross-identifications, auxiliary data
and literature entries in the Simbad database.

In P80 only very few events could be observed, mainly
because ISAAC underwent a long period of maintenance.
In P81 and P83 however the allocated time was completely
filled. 
A total of 4, 125 and 91 events were recorded
in the three periods, respectively. The number was
larger in P81 than in P83, because in this latter
we decided to increase the number of frames recorded
in each event from 7000 to 9500  to optimize
detection.
The brightest and faintest sources among our observed targets
had K=2.45 and K=10.88\,mag respectively, with the median at
K=7.7\,mag.  All the photometric information quoted
in this work comes by default from 2MASS, and 
it should be noted that we have used the K symbol
also for the K$_{\rm s}$ filters employed in 2MASS
as well as in ISAAC.
We also note that in some cases the
source variability 
between the 2MASS epoch and the time 
of our LO measurement could be significant.
This is discussed in Sect.~\ref{sect_unres}.

For all three periods,
the integration time was 3.2~ms. In the burst mode,
this represents also the effective sampling time,
i.e. there are no dead times. Also the frame size
was always the same, namely 32x32 pixels corresponding
to about $4\farcs7$ on the sky.
A broad-band K$_{\rm s}$ filter was always used,
except for the very bright source {\object{2MASS 10093146+0935352}}
when we used a narrow band filter centered at 2.07~$\mu$m.
The transmission curve of these two filters in the cold
was properly accounted for in our data reduction.
Out of the total of 220 events, only 191 have useful data and they are
listed in Table~\ref{lo_complete}. The remaining events were affected by
various problems, the more frequent ones being
mispointing of the telescope or
data recording started too late or too early.

Details about the generation of the light curves
from the data cubes and about the data reduction were given
in Paper~I.
Here we only briefly state that the bulk of the raw
data processing has been done via a specific
pipeline (Fors et al. \cite{fors08}) that performs
automatic mask generation, data extraction, and wavelet-transform
analysis
to derive first guesses of the main free parameters in the fits.
A more detailed interactive analysis was then
carried out on a selected
number of sources using both
model-dependent and model-independent procedures
(ALOR and CAL respectively,
Richichi et al. \cite{richichi96}
and
Richichi \cite{CAL}).
We also used the approach described in
Richichi et al. (\cite{richichi96})
to compute upper limits on the angular sizes of the
unresolved sources.
As already derived in Papers~I-II, the LO technique
at VLT/ISAAC effectively reaches
$\approx 1$ mas in angular resolution and
K$\approx 12$\,mag in sensitivity, thus representing
the most powerful combination presently available
for high angular resolution in the near-IR.

\section{Results}\label{results}
A list of the sources observed, including details of observation
and comments, is provided in Table~\ref{lo_complete}.
They are distributed over about half of the total length of the ecliptic.
There is no overlap with the sources included in Papers~I-II.
While the majority of the targets have
near-IR colors  consistent with those of giant or dwarf stars, 
in several cases more extreme colors are also observed. 
Although the closest distance to the Galactic Center is about 7$\degr$,
a significant part of the sample is located
in the general direction to the galactic bulge and in some
cases reddening may have been significant.
However also a few individual cases of marked extinction probably
of local origin have been noticed, as discussed below.

About one-third of the sources have some cross-identification
in the Simbad database, and less than one-sixth have 
a spectral determination. Even for these, the literature
entries are mostly very scarce.
In 
Tables~\ref{tab:results1}-~\ref{tab:results2}
we list the sources for which we found a positive result, 
either as binary or triple stars or
with a resolved angular diameter.
Unlike Table~\ref{lo_complete} which is ordered by
date and time of occultation, 
Tables~\ref{tab:results1}-~\ref{tab:results2}
are ordered by right ascension
for ease of reference. They
follow the same format of Paper~I and of previous
papers referenced therein.
In summary, the columns list the value of the fitted linear rate of
the event V, its deviation from the predicted rate V$_{\rm{t}}$,
the local lunar limb slope $\psi$, the position and
contact angles, and the signal-to-noise ratio (SNR). 
In Table~\ref{tab:results1}, the
projected separation and the brightness ratio are given,
and in addition also the individual magnitudes obtained
by decomposition of the 2MASS magnitude.
In Table~\ref{tab:results2}, the
the angular diameter $\phi_{\rm UD}$ is reported
for resolved stars, under the assumption of
a uniform stellar disc, or the characteristic size of the shell.
All angular quantities are computed from the fitted
rate of the event.

In the following of this section we focus mainly on
the sources which were found to be resolved and
for which we can provide some context from existing
literature. We then discuss briefly those without
any known cross-identification.
Without showing figures of all the data and their fits,
we present as illustration only an example of a resolved
diameter and of a binary source.
As in Paper~I and II, we provide
at the end some quantitative evaluations derived from
the large body of
the unresolved sources.

\addtocounter{table}{1}
\begin{table*}
\caption{Summary of results: binaries and triples.}
\label{tab:results1}
\centering          
\begin{tabular}{lcrrrrrrrl}
\hline 
\hline 
\multicolumn{1}{c}{(1)}&
\multicolumn{1}{c}{(2)}&
\multicolumn{1}{c}{(3)}&
\multicolumn{1}{c}{(4)}&
\multicolumn{1}{c}{(5)}&
\multicolumn{1}{c}{(6)}&
\multicolumn{1}{c}{(7)}&
\multicolumn{1}{c}{(8)}&
\multicolumn{1}{c}{(9)}&
\multicolumn{1}{c}{(10)}\\
\multicolumn{1}{c}{Source}&
\multicolumn{1}{c}{V (m/ms)}&
\multicolumn{1}{c}{V/V$_{\rm{t}}$--1}&
\multicolumn{1}{c}{$\psi $($\degr$)}&
\multicolumn{1}{c}{PA($\degr$)}&
\multicolumn{1}{c}{CA($\degr$)}&
\multicolumn{1}{c}{SNR}&
\multicolumn{1}{c}{Sep. (mas)}&
\multicolumn{1}{c}{Br. Ratio}&
\multicolumn{1}{c}{Comments}\\
\hline 
08202097+2044517 & 0.5507 & $-$0.6\% & $-$0.5 & 85  & $-$33   & 305.2 & 22.6 $\pm$ 0.1 & 204.0 $\pm$ 1.2 &    K$_1$=3.1, K$_2$=8.9\\
10093146+0935352 & 0.4685 & $-$11.3\% & $-$7.0 & 161  & 34   & 117.8 & 5.6 $\pm$ 0.1 & 32.1 $\pm$ 0.3 &    K$_1$=2.5, K$_2$=6.2\\
10142407+0843549 & 0.4766 & $-$35.9\% & $-$39.1 & 103  & $-$27   & 3.9 & 9.2 $\pm$ 0.5 & 1.4 $\pm$ 0.1 &     A-B: K=10.3, 10.7 \\
10142407+0843549 & 0.4766 & $-$35.9\% & $-$39.1 & 103  & $-$27   & 2.0 & 21.4 $\pm$ 1.2 & 3.6 $\pm$ 0.3 &     A-C: K=10.3, 11.7\\
10323402+0555561 & 0.4967 & 2.4\% & 1.2 & 180  & 50   & 3.5 & 6.8 $\pm$ 0.7 & 1.3 $\pm$ 0.1 &    K$_1$=10.2, K$_2$=10.4\\
10331964+0621594 & 0.3555 & 37.8\% & 8.6 & 70  & $-$60   & 23.7 & 4.1 $\pm$ 0.2 & 6.6 $\pm$ 0.2 &     A-B: K=8.1, 9.5 \\
10331964+0621594 & 0.3555 & 37.8\% & 8.6 & 70  & $-$60   & 22.9 & 8.4 $\pm$ 0.2 & 9.0 $\pm$ 0.1 &     A-C: K=8.1, 10.4 \\
14461088-2110332 & 0.6264 & $-$9.1\% & $-$10.1 & 77  & $-$34   & 135.4 & 141.8 $\pm$ 2.9 & 1.0203 $\pm$ 0.0003 &    K$_1$=5.8, K$_2$=5.8\\
14462093-2128243 & 0.3653 & $-$33.6\% & $-$16.5 & 130  & 30   & 103.9 & 8.8 $\pm$ 0.1 & 48.9 $\pm$ 0.4 &    K$_1$=5.7, K$_2$=10.0\\
17073892-2554521 & 0.4248 & $-$4.8\% & $-$2.6 & 33  & $-$48   & 72.6 & 6.76 $\pm$ 0.03 & 8.1 $\pm$ 0.02 &    K$_1$=5.6, K$_2$=7.9\\
17095865-2549195 & 0.5061 & $-$5.0\% & $-$2.7 & 31  & $-$49   & 72.5 & 6.6 $\pm$ 0.4 & 23.3 $\pm$ 0.3 &    K$_1$=6.7, K$_2$=10.2\\
18353398-2616447 & 0.5386 & $-$3.8\% & $-$2.7 & 27  & $-$41   & 52.6 & 37.5 $\pm$ 0.2 & 26.2 $\pm$ 0.1 &    K$_1$=6.3, K$_2$=9.8\\
19291020-2357229 & 0.3468 & $-$13.3\% & $-$5.2 & 7  & $-$55   & 55.8 & 9.4 $\pm$ 0.3 & 11.0 $\pm$ 0.1 &    K$_1$=7.2, K$_2$=9.8\\
19292834-2421555 & 0.4440 & $-$3.1\% & $-$1.6 & 109  & 47   & 135.6 & 8.8 $\pm$ 0.1 & 21.1 $\pm$ 0.1 &    K$_1$=5.5, K$_2$=8.8\\
19310667-2410011 & 0.5450 & 6.5\% & 3.6 & 109  & 47   & 133.9 & 14.9 $\pm$ 0.1 & 46.9 $\pm$ 0.2 &    K$_1$=4.6, K$_2$=8.7\\
19315937-2404415 & 0.4773 & $-$3.6\% & $-$1.6 & 113  & 50   & 35.7 & 9.8 $\pm$ 0.5 & 19.3 $\pm$ 0.4 &    K$_1$=8.0, K$_2$=11.2\\
20384192-1714580 & 0.7197 & 9.2\% & 18.0 & 91  & 36   & 6.6 & 37.9 $\pm$ 6.6 & 8.5 $\pm$ 0.3 &    K$_1$=10.4, K$_2$=12.8\\
20395807-1659559 & 0.7560 & $-$0.5\% & $-$1.9 & 52  & 6   & 16.3 & 25.1 $\pm$ 2.2 & 9.4 $\pm$ 0.3 &    K$_1$=9.4, K$_2$=11.8\\
20402960-1640247 & 0.6806 & 17.5\% & 10.7 & 23  & $-$35   & 79.2 & 8.9 $\pm$ 0.2 & 34.3 $\pm$ 0.6 &    K$_1$=6.7, K$_2$=10.5\\
20412965-1638169 & 0.7910 & $-$2.7\% & $-$3.7 & 34  & $-$25   & 7.1 & 57.8 $\pm$ 0.5 & 4.4 $\pm$ 0.07 &    K$_1$=9.8, K$_2$=11.4\\
\hline 
\hline 
\end{tabular}
\end{table*}

\begin{table*}
\caption{Summary of results: angular sizes.}
\label{tab:results2}
\centering          
\begin{tabular}{lcrrrrrrrrl}
\hline 
\hline 
\multicolumn{1}{c}{(1)}&
\multicolumn{1}{c}{(2)}&
\multicolumn{1}{c}{(3)}&
\multicolumn{1}{c}{(4)}&
\multicolumn{1}{c}{(5)}&
\multicolumn{1}{c}{(6)}&
\multicolumn{1}{c}{(7)}&
\multicolumn{1}{c}{(8)}&
\multicolumn{1}{c}{(9)}\\
\multicolumn{1}{c}{Source}&
\multicolumn{1}{c}{V (m/ms)}&
\multicolumn{1}{c}{V/V$_{\rm{t}}$--1}&
\multicolumn{1}{c}{$\psi $($\degr$)}&
\multicolumn{1}{c}{PA($\degr$)}&
\multicolumn{1}{c}{CA($\degr$)}&
\multicolumn{1}{c}{SNR}&
\multicolumn{1}{c}{$\phi_{\rm UD}$ (mas)}&
\multicolumn{1}{c}{Comments}\\
\hline 
17151916-2730178 & 0.5433 & $-$3.5\% & $-$13.3 & 85  & $-$2   & 94.2 & 3.10 $\pm$ 0.05 &  \\
17184190-2716075 & 0.6278 & $-$4.2\% & $-$4.5 & 54  & $-$26   & 61.1 & 3.74 $\pm$ 0.02 & extended \\
19590315-2215516 & 0.6080 & $-$0.3\% & $-$0.7 & 43  & $-$16   & 176.6 & 2.35 $\pm$ 0.01 &  \\
\hline 
\hline 
\end{tabular}
\end{table*}

\subsection{Resolved sources with known cross-identifications}\label{resolved}

{\object{08202097+2044517}}: 
this object coincides with
{\object{HR 3264}}={\object{SAO 80112}}, 
a bright K1~III star
with numerous bibliographical entries, 
notably about its radial velocity and rotation.
The former does not seem to vary noticeably, and the
latter is very slow
(Famaey et al.~\cite{famaey2005},
de Medeiros \& Mayor, \cite{medeiros1995}).
This star has well determined photometry
from the visual to the mid-infrared.
Gondoin (\cite{gondoin1999}) estimated 
T$_{\rm eff}$=$4600$\,K, and a mass of 2.7~M$_{\sun}$
The observed (B-V)=1.14\,mag suggests very little 
visual extinction.  
The angular diameter has been estimated at
$1.12\pm0.02$\,mas
(M{\'e}rand et al.~\cite{merand2005})
but we could not resolve it in our observation.
However, we detect for the first time a very faint companion,
with projected separation of 22.6\,mas. Assuming a total
magnitude of 3.13 from 2MASS (which however had to include
a saturation correction) we derive a magnitude
K=8.9 for the companion. Follow-up by long-baseline
interferometry is very challenging due to the magnitude
difference, but AO imaging at a large telescope
might detect the companion.
The use of this star as an interferometric calibrator
as proposed by M{\'e}rand et al.(\cite{merand2005}) should
be considered with care.

{\object{10093146+0935352}}: 
this star is {\object{HD 88071}}, which has several photometric 
determinations from the visual to the near-IR. It is relatively
nearby (Hipparcos parallax $3.50\pm0.85$\,mas) and if one
assumes negligible extinction the colors point to
mid-M giant or M0 supergiant, consistent with the available
spectral type.  It seems to have a 
variable radial velocity, e.g., R$_{\rm v}$=8.06 +/- 0.30 
(Famaey et al.~\cite{famaey2005}), and R$_{\rm v}$=6.1 (Wilson \& Joy \cite{wilson1950}), suggestive of
binarity.   We have discovered for the first time a companion, with projected separation of just
5.6\,mas and 3.8\,mag fainter than the primary in the K band.
Using the nominal Hipparcos distance and our separation,
the companion appears to be at 1.6\,AU
from the primary. In the absence of precise notions on
the spectral class of the primary and due to uncertainties
on the eccentricity, and on the actual radial velocity spread
 and not least to projection effects, any consistency
check between the observed R$_{\rm v}$ differences and
the separation has to be postponed.
We note that the companion is probably beyond the reach of
the largest AO-assisted telescopes, and very challenging
also for long-baseline interferometers.

{\object{10331964+0621594}}: 
this is the G0 star {\object{HD 91411}}. The visual and near-IR 
spectral energy distribution and colors
are consistent with a G dwarf.  Its parallax yields a distance of 143\,pc.  
There is no literature reference listed in Simbad.  Our 
occultation observation revealed a triple system, 
with short projected separations A-B and A-C of 4.1 and 8.4\,mas
respectively. From the total 2MASS magnitude, we derive
K=8.07, K=9.51 and K=10.38\,mag for the three components.
Also this system might be challenging to follow-up
by other techniques.

{\object{14461088-2110332}}:
this well-studied bright star
(V=6.4\,mag, {\object{17 Lib}}, {\object{HR 5504}}, {\object{Fin\,309} })
has an extensive list of measurements and publications which
we summarize here only briefly.
Its spectral type is generally recognized as G2V making it a solar
analog, although F7V was also reported
by Piters et al. (\cite{Piters1998}). The distance is 42\,pc,
and precise parallax and proper motion measurements 
suggest membership in the young-middle age thin disk population
(Bartkevicius \& Gudas \cite{Bart2001}), 
also evidenced in its moderately rich
metallicity 
([Fe/H]=+0.11, Eggen \cite{Eggen98}).

The star is a known binary system with 
almost equal optical brightness components and
a 26\,y period.
Extensive visual and speckle measurements are listed
in the Fourth Interferometric Catalog
(Hartkopf et al. \cite{4IC}, 4IC hereafter)
and, in spite of the orbit
being relatively eccentric $e=0.64$ and not yet completely
observed,  orbital elements are available.
Our LO observation easily resolved the pair,  and in fact
the separation between the two stars (423\,ms, or 265\,m
projected at the lunar limb) was large enough that 
the fringe patterns of the two components had to be
fitted with slightly different limb rate values,  with
a difference of 2$\degr$ in the local limb slope
at the points of contact of the two stars.
Using the average result of the two fits, 
we derive the angular separation listed in Table~\ref{tab:results1},
which is in agreement with the orbital elements provided
in the 4IC (predicted projected separation
149\,mas) when the scatter of the measurements is included
and also considering that other measurements close to our
epoch seem to indicate a smaller-than-predicted separation.
Our flux ratio is very accurately determined,
R=$1.0203 \pm 0.0003$ at K band, and will also help
to disentangle the assignments of the quadrants in
speckle observations which are challenging in the
presence of equal brightness components.
The fact that the companion 
has about the same brightness as the primary in both 
the optical and infrared wavelengths 
shows that this system consists of almost twins. 

{\object{17095865-2549195}}:
this object was noted by Terzan \& Ounnas (\cite{terzan1988}, 
{\object{Terz~V720}}) as a variable star in 
the Sgr~B cloud, based on BVR plates of Palomar Observatory with
R brightness varying between 15.9 and 18\,mag in a few months.  
It has extreme near-infrared colors, with J$=$9.02, 
H$=$7.64, and K$=$6.69\,mag.  
Our lunar occultation observations revealed a companion with
projected separation of just $6.6$\,mas, 3.4\,mag fainter.
It is not clear whether the fainter star in our K band
is the primary or the secondary in the visual, i.e. whether
one of the two stars has much redder colors.
We also note that our observation barely detects a possible
fainter companion at 60\,mas, but this is at the limit
of the noise and we mention it only in case
of findings by future investigations.
Follow-up observations of this object are warranted.  

\begin{figure}
\includegraphics[width=8.8cm]{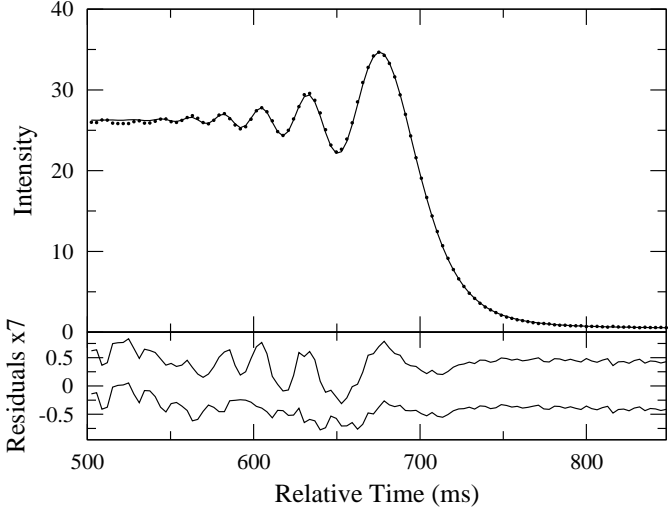}
\caption {
Top panel:
data (dots) and best fit (solid line)
for
{\object{17151916-2730178}}.  The lower panel shows, on an
enlarged scale and
displaced by arbitrary offsets
for clarity, the residuals of the fits by a point-like 
(above, reduced $\chi^2=2.2$) and 
a resolved uniform-disk  model
(below, reduced $\chi^2=1.0$) as listed in
Table~\ref{tab:results2}.
}
\label{fig_diameter}
\end{figure}

{\object{17151916-2730178}}: 
we could record a high SNR light curve for this star,
which yielded a resolved angular diameter of 
$3.02\pm0.04$\,mas (see Fig.~\ref{fig_diameter}). 
No literature reference and no spectral type for this star were
found in the Simbad database, 
but it coincides with {\object{CD-27 11537}} having V=9.6\,mag.
It is however very bright in the near-IR
with K=3.69\,mag. The 2MASS colors (J-H=1.28, 
H-K=0.39\,mag) suggest a highly reddened giant star.  In the
absence of a spectral type 
we assume an early M giant, which would have a diameter $\approx$100~R$_{\sun}$
and absolute K$\approx -4$. Neglecting extinction, it
should thus be at a distance of 340\,pc, thereby subtending 
an angular diameter approximately consistent with our measurement.

{\object{18353398-2616447}}: 
this star coincides with ADS 11463 AB, a relatively
wide binary having a 
separation of $1\farcs6$ along PA=$221\degr$ in 
1927 (Dommanget \& Nys \cite{domm2002}),
and $1\farcs66$ along PA=$218\degr$ at
epoch 1985.36 (Heintz \cite{heintz87}). 
Photometric information includes V=9.9 and 11.3\,mag
for the primary and secondary components, respectively. 
This  star also coincides with
{\object{TYC 6866-75-1}}, whose Tycho magnitudes can be
converted to obtain V=10.15\,mag which however does not
seem in complete agreement with either the sum of the previously
quoted values or with just the primary. 
The difference could be due to the mixed use of visual
and photoelectric magnitude systems.
Our observation reveals a close pair
with a projected separation of $37.5$\,mas along PA=$27\degr$,
and a total flux consistent with the 2MASS K=6.2 magnitude. 
The companion 
is clearly a third (hierarchical) component in the
system, since the AB projected separation should have
been about 40 times larger. However, the issue of whether
this is a companion to ADS 11463~A or B is not completely
clear. We should have seen both A and B in our field of view
and in the range of our light curve, but we do not detect
a second pattern to the limit K$\ga$10\,mag.
In the absence of near-IR photometry, the best guess is that
ADS 11463~A is a late-type giant star (V-K$\approx$3.5\,mag) 
with the close companion detected by us, while B is 
a main sequence star with V-K$\la$1.5\,mag.

{\object{19291020-2357229}}: 
this star coincides with {\object{CD-24 15361}}
and  {\object{TYC 6876-970-1}}, which have no bibliographical
entries. We have detected for the first time a companion
with $9.4$~mas projected separation. The 
K magnitudes of 7.2 and 9.8 for the two components put
follow-up observations at the limit of the capability
of a large long-baseline interferometers such as the
ESO VLTI.

{\object{19310667-2410011}}:
we resolved for the first time this source as binary,
with $14.9$~mas projected separation and
K magnitudes of 4.6 and 8.7 for the two components.
This star coincides with
{\object{IRAS 19280-2416}}  for which only the 
12$\mu$m flux could be measured (0.77~Jy) and for which
no bibliographical entries exist.

{\object{19590315-2215516}}: 
our high-SNR observation permitted us to measure an                              accurate angular diameter of
$2.35$\,mas for this star ({\object{HD~189075}}).
There is no literature 
found in the Simbad database for this K2/K3~III star, but there
is plentiful photometric information for us to
derive 
a reddening correction of E(J-H)=0.132 and
E(H-Ks)=0.088, giving
intrinsic colors of (J-H)$_0$=$0.699$ and (H-Ks)$_0$=$0.13$\,mag 
which are consistent with a~K2/K3III  star
(Bessell \& Brett \cite{bessellbrett}).
Adopting (B-V)$_0$=$1.16$\,mag,
the observed V=9.18\,mag and 
(B-V) of 1.52\,mag, and the standard
value R=A$_{\rm V}$/E(B-V)=3.1, 
we infer A$_{\rm V}$=$1.1$\,mag
which in turn yields a distance of $\approx 330$\,pc.
Our measured angular size hence corresponds to a diameter of 
165~R$_{\sun}$ which, given the 
various parameter uncertainties, 
agrees reasonably with the expected size of the star. 

{\object{20402960-1640247}}: 
this coincides with {\object{BD-17 6053}}, 
which has photometric measurements 
suggesting 
a mid- to late-type dwarf.  There is no Simbad literature reference.  
Our observation revealed for the first time
a companion with projected separation
of 8.9\,mas, almost 4 magnitudes fainter in the K band, and
which will be a difficult target for any other technique.

\subsection{Resolved sources without
 known cross-identifications}\label{resolved-no}
\begin{figure}
\includegraphics[width=8.8cm]{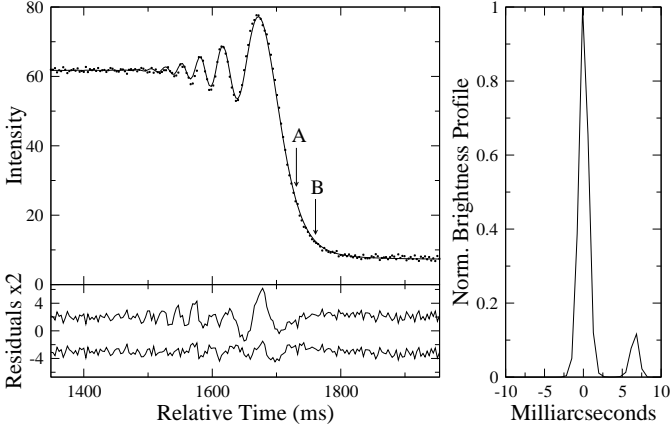}
\caption {{\it Left:} upper panel, data (dots) and best fit (solid line)
for
{\object{2MASS 17073892-2554521}}.  The lower panel shows,
on an enlarged scale and
displaced by arbitrary offsets
for clarity, the residuals of the fits by a point-like 
(above)
and  by a binary star model
(below)
as listed in Table~\ref{tab:results1}.
The normalized $\chi^2$ values are 2.8 and 1.2, respectively.
The times of the geometrical occultation
of two stars are also marked, with their difference
corresponding to a separation of 7\,mas.
{\it Right:} brightness profile reconstructed by the
model-independent CAL method.
}
\label{fig_binary}
\end{figure}

The remaining sources in 
Tables~\ref{tab:results1}-~\ref{tab:results2}
have no known cross-identification in the Simbad
database. They include 8 new binary stars, with
projected separations between 6.8 and 57.8\,mas.
Based on 2MASS magnitude and brightness ratio,
the faintest companion has K=12.7\,mag.
We provide in Fig.~\ref{fig_binary} an example
of the detection of a binary star both by
the model-independent and by the model-dependent
analysis.
We detect {\object{10142407+0843549}} as 
as a first-time triple star, with components
having magnitudes K=10.29, 10.66 and 11.68,
respectively. The projected AB and AC separations
are 9.2 and 21.4\,mas. For this star, like almost
all other binaries, follow-up studies will be
extremely difficult since the combination
of separation and magnitudes are beyond the
current capabilities of either long-baseline
interferometry or AO-assisted single telescopes.

We also detect  {\object{17184190-2716075}} to be
resolved, with a size of 3.74\,mas in the uniform-disk
(UD) sense. The CAL profile for this source shows
faint circumstellar emission extending at least
10\,mas from the central star, possibly asymmetric.
We interpret this as the signature of a circumstellar
shell and therefore the UD diameter becomes
less meaningful. 
There is no spectral classification available,
but we note that the 2MASS colors for this star
are extremely red (J-K=4\,mag) and consistent with
the presence of local extinction. Further studies
are warranted.

\subsection{Unresolved sources and performance}\label{sect_unres}
\begin{figure}
\includegraphics[width=8.8cm]{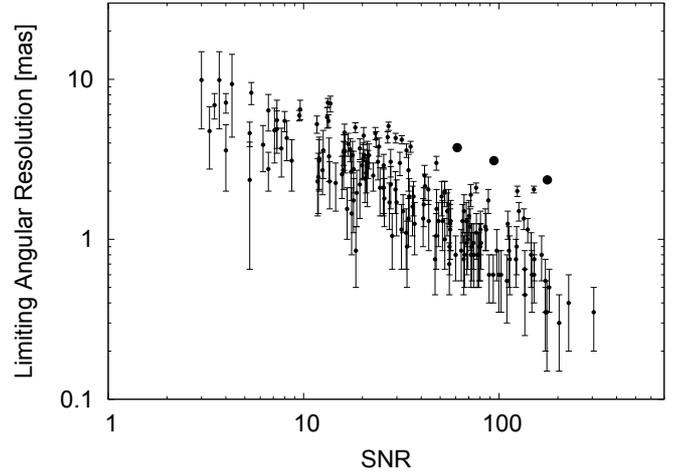}
\caption {Sources for which an upper limit on the angular diameter
could be established using the method already adopted
in Paper~I-II are
marked with the smaller dots and their errorbars. 
The three larger circles are the sources
which we found to be resolved, with errorbars too small
to be visible on this scale.
}
\label{fig_limres}
\end{figure}

We have used the same $\chi^2$-based 
procedure adopted in Papers~I-II to
compute upper limits on the angular size of 177 unresolved
sources. This number includes the individual components
of the binary and triple stars.
In the present paper we provide the corresponding
plot on a logarithmic scale in Fig.~\ref{fig_limres},
confirming a clear dependence of the 
limiting angular resolution on the SNR.
In summary, the angular resolution against
SNR follows a log-log relation, with the
former being $\la$0.6\,mas for SNR$\ge$100. At the faint
end, the angular resolution is $\approx$5\,mas for SNR=10.
The scatter is significant, and is due among other factors
to the dependence of the light curves quality on the
phase of the Moon and on whether and when an active optics
correction had been performed prior to the observations.
A few additional stars resulted to be formally marginally
resolved, but the associated errorbars did not provide
reliable evidence compared to the general scatter of
Fig.~\ref{fig_limres}.
The three sources that we found to be resolved are
also plotted, showing that they lie clearly above the
scatter of the mean relationship.

The measured counts follow a relationship
which is very close to the predicted ISAAC performance
according to its exposure time calculator, as was
discussed in Papers~I-II.
This relationship is maintained from the bright to 
the faint end over about 8.5 magnitudes, with only
a few discrepancies in the direction of fainter fluxes
only, and with mostly well understood reasons
such as occasional very bad image quality due to our need
to stop the active mirror corrections during
LO observations.
This confirms our conclusion stated in Paper~II,
that in the vicinity of the Galactic Center the
main reason for the marked scatter in measured
counts against 2MASS fluxes was probably due
to intrinsic variability of the sources.

The large number of unresolved sources resulting
from our observations, including Papers~I-II, can
be of interest for some applications such as the
search for reliable interferometric calibrators
with accurate upper limits on the size. We intend
to prepare in the near future
such a list, where we will also flag current
interferometric calibrators that we have found to
have companions.

\section{Conclusions}
We have extended the near-IR lunar occultation
observations at the ESO VLT using the ISAAC instrument
in burst mode with high time resolution, first reported
in Papers~I-II. 
In the present paper we have described the first-time
implementation of such observations in service mode
as a filler program. For this,
a large database of predicted
events was prepared, and staff astronomers on the
site would select one whenever a short gap of
time ($\approx 5$\,minutes) was available.

A total of 191 stars were measured in this fashion
between March 2008 and September 2009, leading to
the discovery of 15 new binaries and 2 new
triple stars. An additional star is a well-known
binary, for which we confirm the orbital predictions
and provide a high-accuracy brightness ratio in
the near-IR. 
We also determined for the first time uniform-disk
sizes for three stars, one of which is likely
a star with compact circumstellar emission.

We have confirmed the performance of this technique
already presented in the previous papers. With the
ability to separate binary stars as close as very
few milliarcseconds, with flux ratios exceeding
5 magnitudes, and a limiting sensitivity close
to K=12.5\,mag, the LO method at the ESO VLT
addresses a parameter space for binary stars
not easily available with other techniques.
In spite of the random, fixed-time nature of the
events and thanks to the very economical use
of telescope resources, LO are a competitive
technique especially for serendipitous discoveries
of binary stars, and are ideally suited for
service mode observations.

\begin{acknowledgements}
AR wishes to thank the National Astronomical Research
Institute of Thailand, where he worked on the
data used in this paper, under support from the
\emph{ESO Director General's Discretionary Fund}.
OF is partially supported by 
\emph{MCYT-SEPCYT Plan Nacional I+D+I AYA\#2008-01225}.
We are grateful to Dr. E.~Horch for his valuable
comments on the speckle measurements of
{\object{17 Lib}}.
This research made use of the Simbad database,
operated at the CDS, Strasbourg, France, and
of data products from the Two Micron All Sky Survey, 
which is a joint project of the University of Massachusetts 
and the Infrared Processing and Analysis Center/California Institute 
of Technology, funded by the National Aeronautics and 
Space Administration and the National Science Foundation.
\end{acknowledgements}

\longtab{1}{
\begin{longtable}{lccrrrll}
\caption{\label{lo_complete}
List of the recorded occultation events}\\
\hline\hline
\multicolumn{1}{c}{2MASS id}&
\multicolumn{1}{c}{Date}&
\multicolumn{1}{c}{UT}&
\multicolumn{1}{c}{J}&
\multicolumn{1}{c}{H}&
\multicolumn{1}{c}{K}&
\multicolumn{1}{c}{Sp}&
\multicolumn{1}{c}{Cross-Id} \\
\hline
\endfirsthead
\caption{continued.}\\
\hline\hline
\multicolumn{1}{c}{2MASS id}&
\multicolumn{1}{c}{Date}&
\multicolumn{1}{r}{UT}&
\multicolumn{1}{c}{J}&
\multicolumn{1}{c}{H}&
\multicolumn{1}{c}{K}&
\multicolumn{1}{c}{Sp}&
\multicolumn{1}{c}{Cross-Id} \\
\hline
\endhead
\hline
\endfoot
09101668+1739225 & 18/03/2008 & 4:39:05 & 4.65 & 3.60 & 3.13 &  &  \\
09103221+1738136 & 18/03/2008 & 4:45:45 & 8.46 & 8.20 & 8.11 & G0 & SAO 98413 \\
09553754+1339194 & 18/03/2008 & 23:41:22 & 9.59 & 9.01 & 8.89 &  &  \\
15280539-2403098 & 15/06/2008 & 23:28:00 & 9.69 & 9.23 & 9.09 &  & RAVE J152805.4-240309 \\
15284586-2404001 & 15/06/2008 & 23:42:48 & 7.90 & 7.35 & 7.22 & K0 & SAO 183503 \\
15290862-2341578 & 15/06/2008 & 23:52:25 & 8.45 & 8.18 & 8.09 & G2/G3V & HD 137858 \\
15303449-2417581 & 16/06/2008 & 1:06:15 & 8.83 & 8.48 & 8.41 & G0 & SAO 183530 \\
15320739-2427130 & 16/06/2008 & 2:19:21 & 8.47 & 8.03 & 7.88 &  & CD-24 146 \\
15334356-2426400 & 16/06/2008 & 3:20:14 & 8.29 & 7.73 & 7.59 &  & CD-24 12159 \\
15340268-2413409 & 16/06/2008 & 3:48:48 & 9.69 & 8.87 & 8.64 &  &  \\
15362013-2437453 & 16/06/2008 & 5:24:35 & 9.74 & 9.02 & 8.75 &  &  \\
15363136-2435557 & 16/06/2008 & 5:29:40 & 9.17 & 8.51 & 8.37 &  &  \\
15363969-2435448 & 16/06/2008 & 5:35:00 & 7.96 & 7.29 & 7.04 &  &  \\
15365031-2435115 & 16/06/2008 & 5:41:34 & 7.49 & 6.87 & 6.66 & K0 & CD-24 12195 \\
15374794-2425028 & 16/06/2008 & 6:25:18 & 9.68 & 9.36 & 9.29 &  & SAO 183616 \\
16221656-2624434 & 17/06/2008 & 0:36:25 & 9.40 & 8.37 & 8.12 &  &  \\
16223585-2619365 & 17/06/2008 & 0:41:50 & 10.67 & 9.60 & 9.26 &  &  \\
18282448-2644289 & 12/08/2008 & 23:12:16 & 6.34 & 5.35 & 5.00 &  &  \\
18283098-2701284 & 12/08/2008 & 23:19:16 & 7.03 & 6.07 & 5.70 &  &  \\
18285404-2657280 & 12/08/2008 & 23:28:25 & 9.32 & 8.25 & 7.76 &  &  \\
18290580-2659280 & 12/08/2008 & 23:39:33 & 6.66 & 5.75 & 5.37 &  &  \\
18290708-2702328 & 12/08/2008 & 23:47:47 & 8.58 & 7.73 & 7.15 & M4 & V1890 Sgr \\
18293371-2654429 & 12/08/2008 & 23:54:18 & 8.05 & 7.13 & 6.69 &  &  \\
18284272-2636523 & 13/08/2008 & 0:03:10 & 8.49 & 7.53 & 7.18 &  &  \\
18290429-2637219 & 13/08/2008 & 0:07:38 & 9.48 & 8.49 & 8.08 &  &  \\
18295980-2700098 & 13/08/2008 & 0:21:32 & 7.65 & 6.66 & 6.18 &  &  \\
18300895-2643284 & 13/08/2008 & 0:26:52 & 9.67 & 8.66 & 8.27 &  &  \\
18354857-2632039 & 13/08/2008 & 4:33:09 & 7.22 & 6.24 & 5.86 &  &  \\
18353398-2616447 & 13/08/2008 & 4:45:24 & 7.09 & 6.42 & 6.22 &  & ADS 11463 B \\
18360731-2621282 & 13/08/2008 & 4:51:09 & 9.37 & 8.29 & 7.80 &  &  \\
18364348-2615555 & 13/08/2008 & 5:17:19 & 8.89 & 7.76 & 7.28 &  &  \\
18370713-2621061 & 13/08/2008 & 5:23:44 & 6.82 & 5.80 & 5.31 &  &  \\
18364419-2609382 & 13/08/2008 & 5:31:47 & 6.69 & 5.96 & 5.76 & K & SAO 187038 \\
18371238-2610222 & 13/08/2008 & 5:40:15 & 7.83 & 6.80 & 6.13 & M7 & V2012 Sgr \\
18373700-2608333 & 13/08/2008 & 5:53:43 & 9.01 & 7.83 & 7.20 &  &  \\
18372373-2604267 & 13/08/2008 & 6:00:24 & 9.42 & 8.25 & 7.81 &  &  \\
19221963-2458197 & 13/08/2008 & 23:21:17 & 9.98 & 9.28 & 9.09 &  &  \\
19220718-2435041 & 13/08/2008 & 23:29:36 & 9.15 & 8.44 & 8.23 &  &  \\
19221543-2433187 & 13/08/2008 & 23:44:10 & 9.81 & 9.01 & 8.82 &  &  \\
19233221-2445542 & 14/08/2008 & 0:00:40 & 7.00 & 6.19 & 5.92 &  & CD-25 13997 \\
19234714-2444250 & 14/08/2008 & 0:11:21 & 9.67 & 8.78 & 8.44 &  &  \\
19235395-2451333 & 14/08/2008 & 0:18:40 & 9.75 & 9.22 & 9.09 &  &  \\
19241404-2443562 & 14/08/2008 & 0:30:13 & 8.06 & 7.45 & 7.29 &  & CD-24 15292 \\
19241691-2436572 & 14/08/2008 & 0:39:10 & 8.55 & 7.82 & 7.66 &  &  \\
19245406-2435402 & 14/08/2008 & 1:05:35 & 9.23 & 8.51 & 8.31 &  &  \\
19250663-2433517 & 14/08/2008 & 1:16:54 & 7.18 & 6.40 & 6.19 &  &  \\
19272550-2417011 & 14/08/2008 & 3:18:22 & 7.57 & 6.82 & 6.54 &  &  \\
19275042-2421331 & 14/08/2008 & 3:23:56 & 9.86 & 9.03 & 8.82 &  &  \\
19280328-2420177 & 14/08/2008 & 3:33:55 & 9.46 & 8.60 & 8.36 &  &  \\
19283665-2421189 & 14/08/2008 & 3:55:47 & 6.42 & 5.59 & 5.29 &  & CD-24 15356 \\
19284188-2415497 & 14/08/2008 & 4:04:01 & 8.93 & 8.09 & 7.83 &  &  \\
19280572-2405325 & 14/08/2008 & 4:14:36 & 8.10 & 7.25 & 6.95 &  &  \\
19292074-2415482 & 14/08/2008 & 4:29:21 & 7.25 & 6.50 & 6.31 &  & CD-24 15366 \\
19292834-2421555 & 14/08/2008 & 4:37:00 & 6.61 & 5.74 & 5.47 &  &  \\
19293804-2412464 & 14/08/2008 & 4:42:36 & 9.65 & 8.88 & 8.69 &  &  \\
19284878-2359162 & 14/08/2008 & 4:52:06 & 9.34 & 8.59 & 8.36 &  &  \\
19291020-2357229 & 14/08/2008 & 5:03:41 & 7.76 & 7.25 & 7.15 &  & CD-24 15361 \\
19301024-2404280 & 14/08/2008 & 5:09:42 & 6.90 & 6.12 & 5.83 &  & CD-24 15375 \\
19301492-2418333 & 14/08/2008 & 5:14:46 & 9.86 & 9.12 & 8.92 &  &  \\
19302217-2358275 & 14/08/2008 & 5:24:53 & 7.97 & 7.19 & 6.93 &  &  \\
19304386-2359422 & 14/08/2008 & 5:33:29 & 9.63 & 8.76 & 8.42 &  &  \\
19310667-2410011 & 14/08/2008 & 5:45:20 & 5.78 & 4.89 & 4.54 &  & IRAS 19280-2416 \\
19312730-2352517 & 14/08/2008 & 6:03:49 & 7.55 & 6.82 & 6.63 &  & TYC 6876-820-1 \\
19315064-2405364 & 14/08/2008 & 6:14:32 & 9.85 & 9.26 & 9.13 &  &  \\
19315937-2404415 & 14/08/2008 & 6:20:06 & 8.97 & 8.19 & 7.95 &  &  \\
19322078-2351126 & 14/08/2008 & 6:31:35 & 9.42 & 9.13 & 9.06 & F0 & V440 Sgr \\
19323909-2348232 & 14/08/2008 & 6:42:57 & 8.22 & 7.38 & 7.16 &  &  \\
19324040-2401451 & 14/08/2008 & 6:48:37 & 9.64 & 8.86 & 8.69 &  &  \\
19330835-2355321 & 14/08/2008 & 6:59:13 & 9.69 & 8.87 & 8.60 &  &  \\
20153095-2127522 & 14/08/2008 & 23:47:19 & 6.79 & 6.27 & 6.16 & G9IV & SAO 189073 \\
20153726-2121357 & 14/08/2008 & 23:57:45 & 8.13 & 7.55 & 7.38 &  & BD-21 5661 \\
20160387-2122027 & 15/08/2008 & 0:12:06 & 8.44 & 7.80 & 7.66 &  & BD-21 5666 \\
17141098-2720526 & 07/09/2008 & 23:19:39 & 9.22 & 8.07 & 7.43 &  & Terz V 879 \\
17142802-2722488 & 07/09/2008 & 23:26:57 & 10.38 & 9.26 & 8.67 &  & Terz V 891 \\
17143289-2741323 & 07/09/2008 & 23:33:15 & 8.18 & 7.09 & 6.55 &  &  \\
17141958-2718281 & 07/09/2008 & 23:37:43 & 9.64 & 7.87 & 6.79 &  &  \\
17150142-2731588 & 07/09/2008 & 23:42:34 & 9.53 & 7.86 & 6.87 &  & Terz V 923 \\
17150849-2734106 & 07/09/2008 & 23:48:45 & 12.50 & 9.98 & 8.64 &  &  \\
17151916-2730178 & 07/09/2008 & 23:56:18 & 5.36 & 4.08 & 3.69 &  & CD-27 11537 \\
17151713-2739202 & 08/09/2008 & 0:02:23 & 9.31 & 7.84 & 7.10 &  &  \\
17145454-2717035 & 08/09/2008 & 0:11:36 & 11.22 & 9.58 & 8.84 &  &  \\
17154469-2726100 & 08/09/2008 & 0:18:09 & 8.79 & 7.37 & 6.30 &  & Terz V 956 \\
17155252-2737047 & 08/09/2008 & 0:25:41 & 8.54 & 7.36 & 6.77 &  & Terz V 963 \\
17160728-2729216 & 08/09/2008 & 0:32:16 & 10.41 & 9.01 & 8.46 &  &  \\
17161161-2727347 & 08/09/2008 & 0:36:13 & 9.73 & 8.15 & 7.37 &  &  \\
17161831-2732513 & 08/09/2008 & 0:40:44 & 9.63 & 8.23 & 7.62 &  &  \\
17155764-2718013 & 08/09/2008 & 0:45:39 & 9.67 & 8.30 & 7.75 &  &  \\
17164441-2732526 & 08/09/2008 & 0:59:46 & 9.82 & 8.31 & 7.47 &  &  \\
17165182-2730071 & 08/09/2008 & 1:03:56 & 8.63 & 7.35 & 6.79 &  &  \\
17165491-2724095 & 08/09/2008 & 1:08:54 & 6.79 & 5.62 & 5.18 &  &  \\
17170627-2725094 & 08/09/2008 & 1:15:27 & 9.09 & 7.68 & 6.95 &  &  \\
17171581-2732238 & 08/09/2008 & 1:21:48 & 9.76 & 8.50 & 7.89 &  &  \\
17170845-2738355 & 08/09/2008 & 1:27:33 & 6.57 & 5.40 & 4.81 &  &  \\
17181162-2719246 & 08/09/2008 & 2:02:04 & 9.87 & 8.43 & 7.57 &  &  \\
17180561-2738288 & 08/09/2008 & 2:13:31 & 10.00 & 8.39 & 7.78 &  & Terz V 1063 \\
17184512-2728255 & 08/09/2008 & 2:18:46 & 8.77 & 7.65 & 7.12 &  &  \\
17184190-2716075 & 08/09/2008 & 2:23:57 & 11.41 & 9.05 & 7.37 &  &  \\
17190268-2725134 & 08/09/2008 & 2:28:31 & 8.62 & 7.45 & 6.67 &  & Terz V 1096 \\
17190776-2718024 & 08/09/2008 & 2:35:06 & 7.84 & 6.63 & 5.87 &  & Terz V 1104 \\
17185378-2736499 & 08/09/2008 & 2:45:05 & 8.56 & 7.38 & 6.87 &  &  \\
19024623-2550139 & 09/09/2008 & 23:33:09 & 8.43 & 7.49 & 7.21 &  &  \\
19024174-2526274 & 09/09/2008 & 23:46:37 & 9.33 & 8.32 & 8.04 &  &  \\
19031126-2549091 & 09/09/2008 & 23:54:24 & 9.63 & 8.86 & 8.62 &  &  \\
19033600-2541322 & 10/09/2008 & 0:02:07 & 7.42 & 6.44 & 6.05 &  &  \\
19031235-2523586 & 10/09/2008 & 0:13:09 & 8.52 & 7.69 & 7.42 &  &  \\
19030489-2521534 & 10/09/2008 & 0:22:13 & 8.53 & 7.60 & 7.33 &  &  \\
19072657-2500098 & 10/09/2008 & 3:23:18 & 9.68 & 8.79 & 8.50 &  &  \\
19071316-2457111 & 10/09/2008 & 3:31:41 & 9.89 & 8.95 & 8.64 &  &  \\
19075933-2458036 & 10/09/2008 & 3:40:53 & 9.67 & 9.11 & 8.96 &  &  \\
19083084-2503534 & 10/09/2008 & 3:45:14 & 9.19 & 8.40 & 8.20 &  &  \\
19084001-2515467 & 10/09/2008 & 3:52:06 & 6.09 & 5.24 & 5.02 & K2III & HD 178198 \\
19091620-2506556 & 10/09/2008 & 4:10:06 & 6.62 & 5.97 & 5.81 & K2III & HD 178343 \\
19581190-2234268 & 11/09/2008 & 1:31:12 & 9.52 & 8.67 & 8.51 &  &  \\
19582624-2235336 & 11/09/2008 & 1:42:37 & 7.51 & 6.97 & 6.83 & G9IV/V & HD 188941 \\
19584057-2228469 & 11/09/2008 & 1:56:14 & 9.71 & 9.01 & 8.83 &  &  \\
19584940-2233253 & 11/09/2008 & 2:01:58 & 9.73 & 9.23 & 9.11 &  &  \\
19590076-2233239 & 11/09/2008 & 2:12:13 & 9.13 & 8.41 & 8.12 &  &  \\
19583385-2215070 & 11/09/2008 & 2:19:33 & 5.93 & 5.32 & 5.17 & K0III & SAO 188772 \\
19590315-2215516 & 11/09/2008 & 2:30:08 & 6.23 & 5.39 & 5.18 & K2/K3III & HD 189075 \\
08200204+2039484 & 03/04/2009 & 23:33:11 & 9.93 & 9.35 & 9.20 &  &  \\
08202097+2044517 & 03/04/2009 & 23:43:53 & 3.89 & 3.38 & 3.13 & K1III & HR 3264 \\
08201514+2019535 & 04/04/2009 & 0:24:00 & 9.56 & 9.35 & 9.29 &  & TYC 1386-551-1 \\
09221156+1435272 & 05/04/2009 & 4:03:07 & 9.18 & 8.63 & 8.49 &  &  \\
10093146+0935352 & 06/04/2009 & 0:23:23 & 3.62 & 2.66 & 2.45 & M0 & HD88071 \\
10102628+0938503 & 06/04/2009 & 0:35:47 & 10.38 & 10.17 & 10.13 &  &  \\
10105208+0948219 & 06/04/2009 & 0:46:42 & 10.15 & 9.43 & 9.33 &  &  \\
10142407+0843549 & 06/04/2009 & 3:59:13 & 10.17 & 9.65 & 9.54 &  &  \\
10142834+0837110 & 06/04/2009 & 4:15:12 & 8.68 & 8.45 & 8.41 & F8 & SAO 118177 \\
10312220+0635587 & 30/05/2009 & 22:36:08 & 8.96 & 8.44 & 8.36 & K0 & SAO 118346 \\
10320353+0629543 & 30/05/2009 & 23:09:39 & 9.74 & 9.46 & 9.37 &  & TYC 259-1407-1 \\
10314505+0617493 & 30/05/2009 & 23:15:45 & 10.91 & 10.48 & 10.39 &  &  \\
10312381+0612440 & 30/05/2009 & 23:23:40 & 10.79 & 10.13 & 9.94 &  &  \\
10323297+0630074 & 30/05/2009 & 23:36:25 & 10.51 & 9.90 & 9.54 &  & NLTT 24622 \\
10331964+0621594 & 31/05/2009 & 0:19:18 & 8.10 & 7.78 & 7.72 & G0 & HD 91411 \\
10323402+0555561 & 31/05/2009 & 0:27:27 & 10.78 & 10.27 & 10.15 &  &  \\
14450972-2126415 & 26/08/2009 & 0:45:34 & 8.48 & 7.82 & 7.62 &  & TYC 6166-205-1 \\
14461088-2110332 & 26/08/2009 & 1:00:36 & 5.37 & 5.14 & 5.00 & G2V & SAO 182858 \\
14462116-2108358 & 26/08/2009 & 1:08:41 & 6.13 & 5.22 & 4.92 & K3V & HD 129991 \\
14462093-2128243 & 26/08/2009 & 1:22:16 & 6.91 & 6.02 & 5.71 &  &  \\
15184024-2257104 & 23/09/2009 & 0:25:32 & 10.11 & 9.59 & 9.43 &  &  \\
15190984-2257010 & 23/09/2009 & 0:40:39 & 9.70 & 9.06 & 8.92 &  & RAVE J151909.9-225701 \\
17073379-2602224 & 24/09/2009 & 23:34:36 & 9.41 & 8.44 & 8.06 &  &  \\
17073939-2617059 & 24/09/2009 & 23:45:50 & 8.79 & 8.01 & 7.71 &  &  \\
17073892-2554521 & 24/09/2009 & 23:53:47 & 6.98 & 5.95 & 5.46 &  &  \\
17075861-2617544 & 25/09/2009 & 0:03:20 & 7.75 & 6.98 & 6.76 &  &  \\
17083472-2609422 & 25/09/2009 & 0:13:47 & 9.15 & 8.14 & 7.65 &  &  \\
17083267-2553363 & 25/09/2009 & 0:27:26 & 8.52 & 7.53 & 7.08 &  &  \\
17092644-2604470 & 25/09/2009 & 0:45:26 & 8.24 & 7.38 & 7.12 &  &  \\
17093337-2556391 & 25/09/2009 & 0:54:31 & 7.19 & 6.21 & 5.82 &  &  \\
17100320-2556378 & 25/09/2009 & 1:10:57 & 8.21 & 7.24 & 6.60 &  & Terz V 724 \\
17095865-2549195 & 25/09/2009 & 1:22:19 & 9.02 & 7.64 & 6.69 &  & Terz V 720 \\
17102817-2547542 & 25/09/2009 & 1:39:36 & 13.15 & 10.48 & 8.11 &  &  \\
17111258-2558126 & 25/09/2009 & 1:47:24 & 8.57 & 7.50 & 6.88 &  &  \\
17113166-2559346 & 25/09/2009 & 1:57:29 & 5.10 & 4.07 & 3.88 &  & CD-25 12008 \\
17113416-2549111 & 25/09/2009 & 2:06:26 & 8.12 & 7.08 & 6.54 &  &  \\
17112858-2544157 & 25/09/2009 & 2:15:25 & 7.45 & 6.40 & 5.92 &  &  \\
17115791-2545246 & 25/09/2009 & 2:24:00 & 9.16 & 8.10 & 7.67 &  &  \\
19453075-2138258 & 27/09/2009 & 23:07:38 & 8.32 & 7.32 & 7.00 &  &  \\
20343615-1756212 & 28/09/2009 & 23:23:40 & 8.03 & 7.50 & 7.34 &  & BD-18 5712 \\
20345853-1751096 & 28/09/2009 & 23:45:28 & 9.65 & 8.95 & 8.79 &  &  \\
20350747-1755399 & 28/09/2009 & 23:51:16 & 8.74 & 8.07 & 7.90 &  &  \\
20352434-1747177 & 29/09/2009 & 0:09:21 & 7.70 & 7.58 & 7.55 & F1V & SAO 163707 \\
20352929-1745326 & 29/09/2009 & 0:15:05 & 8.41 & 7.68 & 7.52 &  & PPM 722024             \\
20360206-1739338 & 29/09/2009 & 0:46:55 & 11.03 & 10.71 & 10.61 &  &  \\
20360492-1734439 & 29/09/2009 & 0:57:02 & 10.49 & 9.90 & 9.78 &  &  \\
20362350-1739254 & 29/09/2009 & 1:03:22 & 6.98 & 6.14 & 5.85 &  &  \\
20354654-1726398 & 29/09/2009 & 1:12:46 & 11.53 & 11.03 & 10.88 &  &  \\
20363206-1731515 & 29/09/2009 & 1:19:25 & 8.56 & 8.07 & 7.89 &  & BD-18 5722 \\
20355250-1723143 & 29/09/2009 & 1:29:51 & 10.90 & 10.42 & 10.28 &  &  \\
20370756-1741511 & 29/09/2009 & 1:49:34 & 8.78 & 8.22 & 8.07 &  &  \\
20364600-1718497 & 29/09/2009 & 1:59:18 & 10.31 & 9.54 & 9.35 &  &  \\
20361895-1716420 & 29/09/2009 & 2:06:35 & 9.09 & 8.20 & 7.99 &  &  \\
20375963-1722596 & 29/09/2009 & 2:30:43 & 9.87 & 9.41 & 9.34 &  &  \\
20380980-1730064 & 29/09/2009 & 2:41:29 & 9.27 & 9.21 & 9.17 & Ap... & AW Cap \\
20374892-1707208 & 29/09/2009 & 2:56:28 & 7.05 & 7.08 & 7.08 & B9V & SAO 163746 \\
20384192-1714580 & 29/09/2009 & 3:07:00 & 11.04 & 10.45 & 10.31 &  &  \\
20381796-1702511 & 29/09/2009 & 3:18:54 & 9.24 & 8.79 & 8.69 &  & TYC 6334-955-1 \\
20384246-1703552 & 29/09/2009 & 3:24:39 & 5.14 & 4.55 & 4.22 & K4/K5III & HD 196557 \\
20391028-1702091 & 29/09/2009 & 3:40:35 & 9.57 & 8.79 & 8.63 &  &  \\
20385707-1655235 & 29/09/2009 & 3:52:32 & 10.06 & 9.77 & 9.68 &  &  \\
20393924-1657314 & 29/09/2009 & 4:02:13 & 9.61 & 8.94 & 8.80 &  &  \\
20395807-1659559 & 29/09/2009 & 4:08:13 & 10.08 & 9.38 & 9.25 &  &  \\
20400160-1656267 & 29/09/2009 & 4:14:30 & 9.66 & 9.40 & 9.34 &  & BD-17 6050 \\
20402190-1702230 & 29/09/2009 & 4:21:00 & 6.55 & 5.69 & 5.42 &  & BD-17 6051 \\
20403208-1703280 & 29/09/2009 & 4:28:11 & 6.67 & 5.83 & 5.41 & M4 & TX Cap \\
20404462-1654461 & 29/09/2009 & 4:38:56 & 7.50 & 6.86 & 6.71 & K1III & HD 196872 \\
20403343-1646457 & 29/09/2009 & 4:44:39 & 11.52 & 10.93 & 10.82 &  &  \\
20405869-1704068 & 29/09/2009 & 4:51:23 & 7.40 & 6.80 & 6.51 &  & BD-17 6056 \\
20402960-1640247 & 29/09/2009 & 4:59:29 & 7.54 & 6.84 & 6.63 &  & BD-17 6053 \\
20411686-1641389 & 29/09/2009 & 5:10:39 & 6.52 & 5.75 & 5.53 & K2/K3III & SAO 163796 \\
20412965-1638169 & 29/09/2009 & 5:21:27 & 10.29 & 9.73 & 9.59 &  &  \\
\end{longtable}
}
\end{document}